\documentclass[a4paper,11pt]{article}
\usepackage{pos}
\usepackage{slashed}

\title{Investigating SU(3) with $N_f=8$ fundamental fermions at strong renormalized coupling}

\author[a]{Anna Hasenfratz}
\author*[b]{Oliver Witzel}
\onbehalf{\newline Lattice Strong Dynamics collaboration}
\affiliation[a]{Department of Physics, University of Colorado, Boulder, Colorado 80309, USA}

\affiliation[b]{Center for Particle Physics Siegen, Theoretische Physik 1,
Naturwissenschaftlich-Technische Fakultät, Universität Siegen, 57068 Siegen, Germany}

\emailAdd{anna.hasenfratz@colorado.edu}
\emailAdd{oliver.witzel@uni-siegen.de}

\abstract{Lattice simulations have observed a novel strong coupling symmetric mass generation (SMG) phase   for the SU(3) gauge system with $N_f=8$ fundamental fermions (represented by two sets of staggered fields) at very large renormalized coupling ($g^2_{GF} \gtrsim 25$). The results of Phys.~Rev.~D 106 (2022) 014513 suggest that the SMG phase is separated  from the weak coupling, conformal phase by a continuous phase transition, implying that the SMG phase exists in the continuum limit.  To scrutinize these findings, we are generating a set of large volume zero temperature ensembles using nHYP improved staggered fermions with additional Pauli-Villars fields to tame gauge field fluctuations. We consider the low-lying meson spectrum and verify the existence of the SMG phase. Based on a finite size scaling analysis we predict that the phase transition between the strong and weak coupling phases is likely governed by a merged fixed point that is ultraviolet in the strong coupling but infrared in the weak coupling side. This finding suggests that the SU(3) 8-flavor system sits at the opening of the conformal window.}

\FullConference{The 41st International Symposium on Lattice Field Theory (LATTICE2024)\\
 28 July - 3 August 2024\\
Liverpool, UK\\}

\begin{document}
\maketitle

\section{Introduction}
Despite substantial efforts by different groups \cite{Appelquist:2007hu,Deuzeman:2008sc,Fodor:2011tu,Appelquist:2011dp,Aoki:2013xza,Lombardo:2014pda,Aoki:2014oha,Hasenfratz:2014rna,Fodor:2015baa,Aoki:2016wnc,Fodor:2017gtj,Hasenfratz:2017qyr,Appelquist:2018yqe,Chiu:2018edw,Hasenfratz:2019dpr,Fodor:2019ypi,Witzel:2019jbe,Hasenfratz:2020ess,Appelquist:2020xua,Hasenfratz:2022yws,Hasenfratz:2022zsa,LSD:2023uzj,LatticeStrongDynamics:2023bqp}, identifying the onset and nature of the conformal window for SU(3) gauge theories with fermions in the fundamental representation is still an open question.  After establishing the existence of an infrared fixed point (IRFP) for SU(3) gauge with $N_f=10$ and 12 fundamental fermions \cite{Hasenfratz:2023wbr,Hasenfratz:2024fad}, $N_f=8$ moved into the focus and, indeed, small scale simulations reveal interesting phenomena at very strong renormalized coupling ($g^2>25$) \cite{Hasenfratz:2022qan}. The findings in Ref.~\cite{Hasenfratz:2022qan} suggest that $N_f=8$ could be the onset of the conformal window and, moreover, gives rise to a new phase referred to as \emph{symmetric mass generation}\/ (SMG) \cite{Fidkowski:2009dba,You:2014oaa,Ayyar:2014eua,Ayyar:2015lrd,Wang:2022ucy,Catterall:2023nww}. The SMG phase is characterized by being chirally symmetric but  nevertheless features confinement and massive hadrons even in the $am_f\to 0$ chiral limit.

The existence of an SMG phase is strongly connected to anomalies. {}'t Hooft anomaly matching requires that systems with non-vanishing anomalies are either conformal, or exhibit spontaneous symmetry breaking  with massless Goldstone bosons in the infrared. 4-dimensional gauge systems with fundamental fermions must contain multiple of 16 Majorana (or 8 massless Dirac) fermions to be anomaly-free. Thus the minimum number of flavors required for an SMG phase in the continuum is   four with SU(2)  and  eight with SU(3) gauge. Using staggered lattice fermions Ref.~\cite{Butt:2024kxi} has established that the former system indeed has an SMG phase in the continuum and the results also suggest that the system is the opening of the conformal window. Our goal in this work is to investigate similar claims of Ref.~\cite{Hasenfratz:2022qan} for the SU(3) gauge system with  eight staggered lattice fermions.

Reaching such strong renormalized coupling is challenging because many standard actions run into a bulk phase transition (see e.g.~the discussion in Ref.~\cite{Hasenfratz:2022zsa}) when the bare gauge coupling is increased. To reach sufficiently strong renormalized coupling, lattice artifacts need to be kept under control. One possibility is to amend the simulated action by adding massive, bosonic Pauli-Villars (PV) fields which help to keep lattice artifacts under control but due to their large mass are automatically integrated out in the continuum \cite{Hasenfratz:2021zsl}.

In order to investigate and scrutinize the scenario that $N_f=8$ corresponds to the onset  of the conformal window and that a continuum SMG phase exists, we aim to carry out large scale lattice simulations at strong coupling. Building upon the findings of Ref.~\cite{Hasenfratz:2022qan}, we study $N_f=8$ using \emph{massless} nHYP-smeared staggered fermions \cite{Hasenfratz:2007rf} with fundamental and adjoint plaquette gauge action with $\beta_\text{adj}=-\beta_b/4$ where $\beta_b$ denotes the coupling of the fundamental plaquette term. We  further include eight bosonic PV fields with $am_{PV} = 0.75$ for every fermion field. While this work places the focus on generating zero temperature configurations to study the hadronic spectrum, the same action is also used in a companion project by Hasenfratz and Peterson targeting the determination of the renormalization group (RG) $\beta$-function using symmetric $(L/a)^4$ lattices \cite{nf8betafn}. In Tab.~\ref{Tab.ensembles} we list the set of bare gauge couplings $\beta_b$ simulated on $16^3\times 32$, $24^3\times 64$, and $32^3\times 64$ volumes. 

\begin{table}
  \centering
  \begin{tabular}{cc}
    \hline\hline
    $\left(L/a\right)^3\times T/a$ & $\beta_b$\\
    \hline
    $16^3 \times 32$ & 8.30, 8.40, 8.50, 8.52, 8.55, 8.57, 8.60, 8.62, 8.65, \\
                     & 8.67, 8.70, 8.80, 8.90, 9.00, 9.10, 9.20, 9.30, 9.50\phantom{,} \\
    $24^3 \times 64$ & 8.30, 8.40, 8.50, 8.52, 8.55, 8.57, 8.60, 8.62, 8.65, \\
                     & 8.67, 8.70, 8.80, 8.90, 9.00, 9.10, 9.20, 9.30, 9.50\phantom{,} \\
    $32^3 \times 64$ &\phantom{ 8.30, }8.40, 8.50, 8.52, 8.55, 8.57, 8.60, 8.62, 8.65, \\
                     & 8.67, 8.70, 8.80, 8.90, 9.00, 9.10, 9.20, 9.30, 9.50\phantom{,} \\
    \hline\hline
  \end{tabular}
  \caption{Analyzed $N_f=8$ zero mass gauge field ensembles with additional PV fields.}
  \label{Tab.ensembles}
\end{table}

\section{Results}
\begin{figure}[t]
  \centering
  \includegraphics[width=0.49\textwidth]{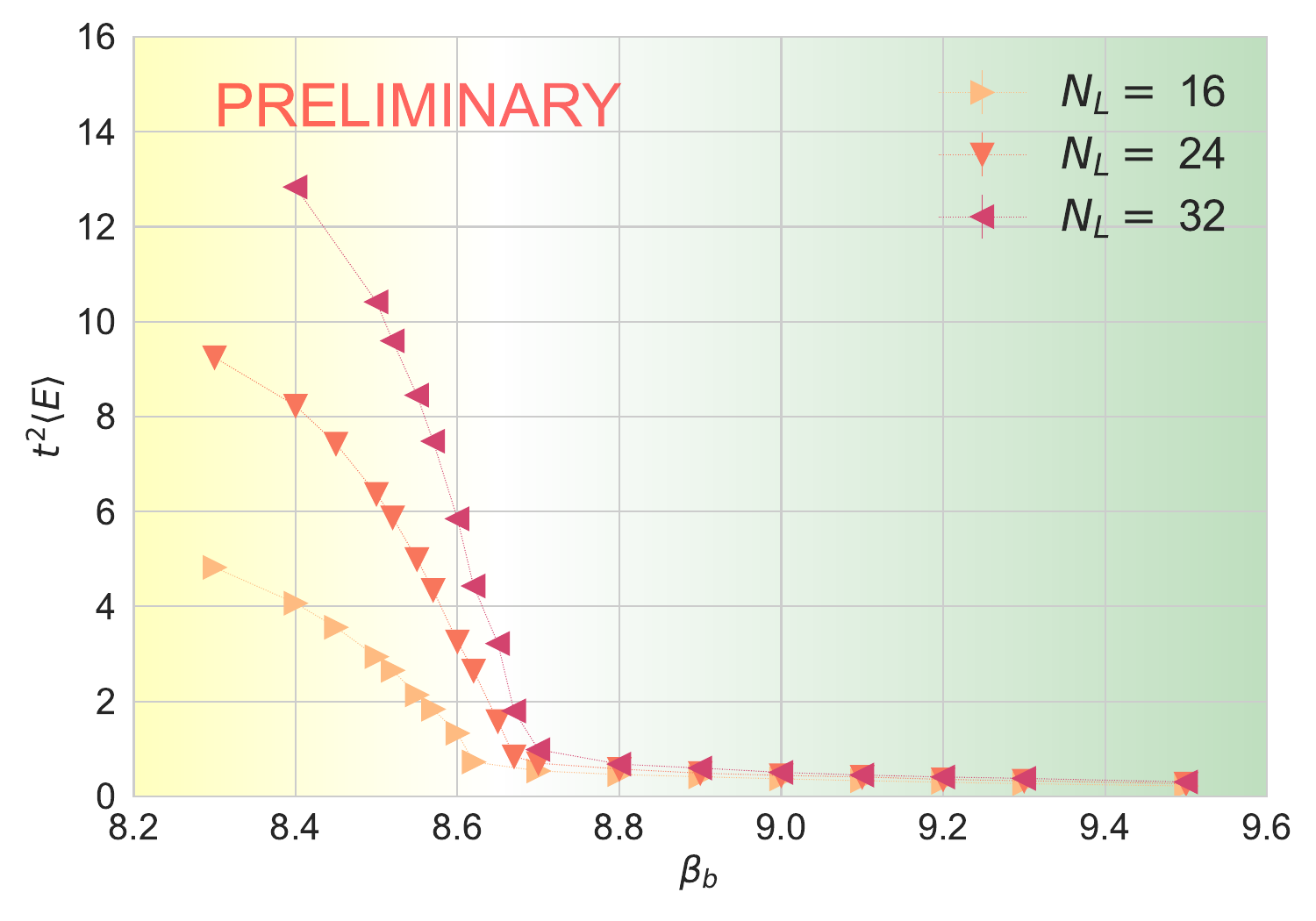}
  \includegraphics[width=0.49\textwidth]{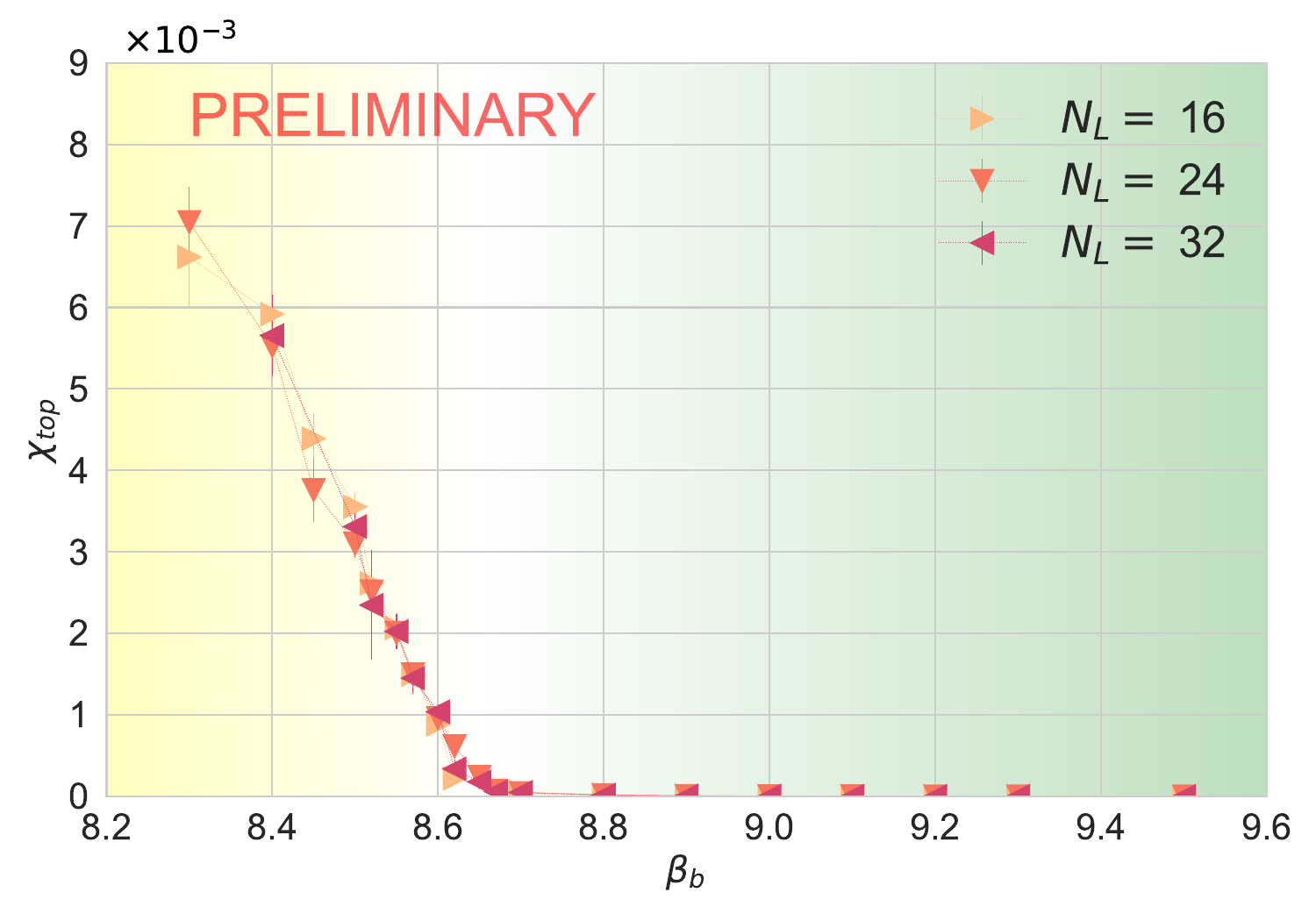}
  \caption{Taking a first qualitative look at our simulations on three different volumes. Left panel: $t^2\langle E \rangle$ at flow time $t=N_L^2/32$ vs.~the bare coupling $\beta_b$ for different spatial lattice volumes with $N_L=L/a$. This quantity is proportional to the gradient flow coupling  $g_c^2$ with $c=0.5$. 
  Right panel: the same as the left panel but for the topological susceptibility, $\chi_\text{top}$.}
  \label{Fig.t2E-beta}
\end{figure}

Performing simulations from weak coupling at $\beta_b=9.50$ to our strongest coupling at $\beta_b=8.30$, we observe that our simulations cross a phase transition. Using e.g.~gradient flow (GF) to determine the finite volume renormalized coupling \cite{Narayanan:2006rf,Luscher:2010iy, Fodor:2012td}, we observe a sudden rise of $g^2$ around $\beta_b\sim 8.65$ as is shown in the left panel of Fig.~\ref{Fig.t2E-beta}. The right panel of Fig.~\ref{Fig.t2E-beta} shows the topological susceptibility $\chi_\text{top} = \left( \langle Q^2 \rangle - \langle Q \rangle^2\right)/V$ where the topological charge is predicted on the gradient flowed configurations at flow time $t=N_L^2/32$,  $N_L=L/a$. Since our simulations are  performed in the chiral limit, in conformal or QCD-like system $Q$ must be zero on every configuration. The GF transformation occasionally inflates a small instanton/dislocation, leading to $Q \ne 0$, but the effect is typically small in magnitude \cite{Hasenfratz:2020vta}.  The sudden and significant rise of $\chi_\text{top}$ around $\beta_b\sim 8.65$ indicates that the vacuum structure in the strong coupling region is very different from the vacuum of a conformal or chirally broken QCD-like system.  Preliminary investigations indicate that a short-range  ordering of the gauge field, referred to as ``shift symmetry breaking'' or $\slashed{S}^4$ in \cite{Cheng:2011ic} masks instantons, thus  they do not lead to zero modes for the fermionic Dirac operator. Gradient flow smoothes the gauge fields and removes this local ordering, reveling the underlying topological structure.  
Furthermore, we notice that simulations near $\beta_b\sim 8.65$ are substantially more expensive than at significantly smaller or larger values of the bare coupling, consistent with critical slowing down expected at a phase transition.

 Calculating the masses of several meson states, we observe \emph{parity doubling}\/ for all measurements, independent of the bare coupling $\beta_b$. Not only the masses, but already the correlators are degenerate configuration by configuration i.e., e.g.~pseudocalar and scalar state correlators are degenerate as are vector and axial states. As expected, at weak coupling the different staggered tastes are nearly degenerate, but taste-splitting opens up as we move to strong coupling. To explore the nature of the phases and the observed phase transition, we focus on the lowest lying pseudoscalar (PS)  and vector meson (V) states.  In Fig.~\ref{Fig.LM-vs-beta} we show the lattice masses $aM_\text{PS}$ and $aM_\text{V}$ as the function of the bare gauge coupling $\beta_b$. For weak coupling (large $\beta_b$) the hadronic masses exhibit a strong volume dependence.  At the same bare coupling $\beta_b$ masses on a smaller volume are larger than on a larger volume. This volume dependence disappears around $\beta_b\sim 8.65$. For stronger coupling the hadronic masses are largely independent of the volume and only vary with bare coupling $\beta_b$. The value of this transition itself also exhibits a small volume dependence.
\begin{figure}[t]
  \includegraphics[width=0.49\textwidth]{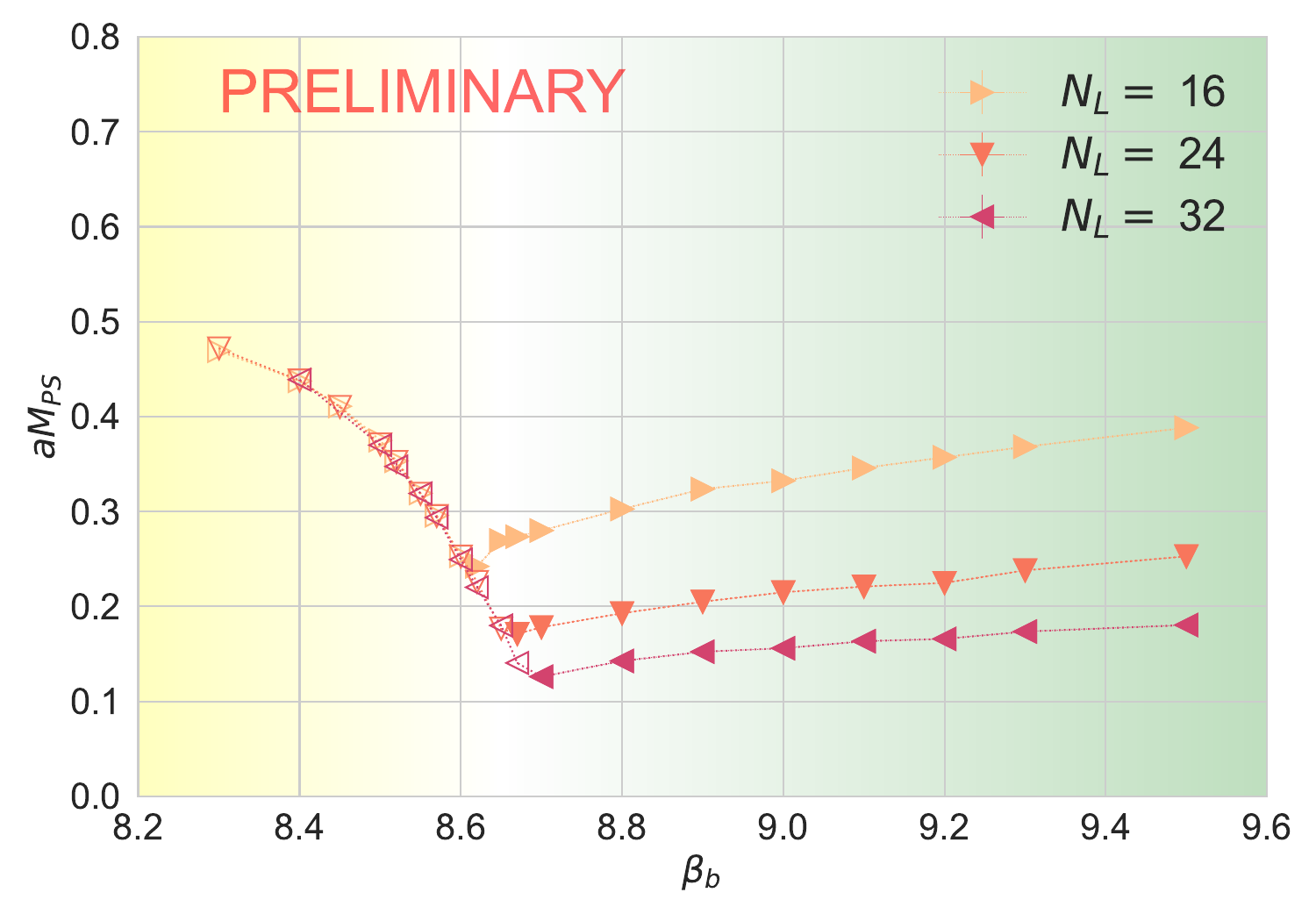}\hfill
  \includegraphics[width=0.49\textwidth]{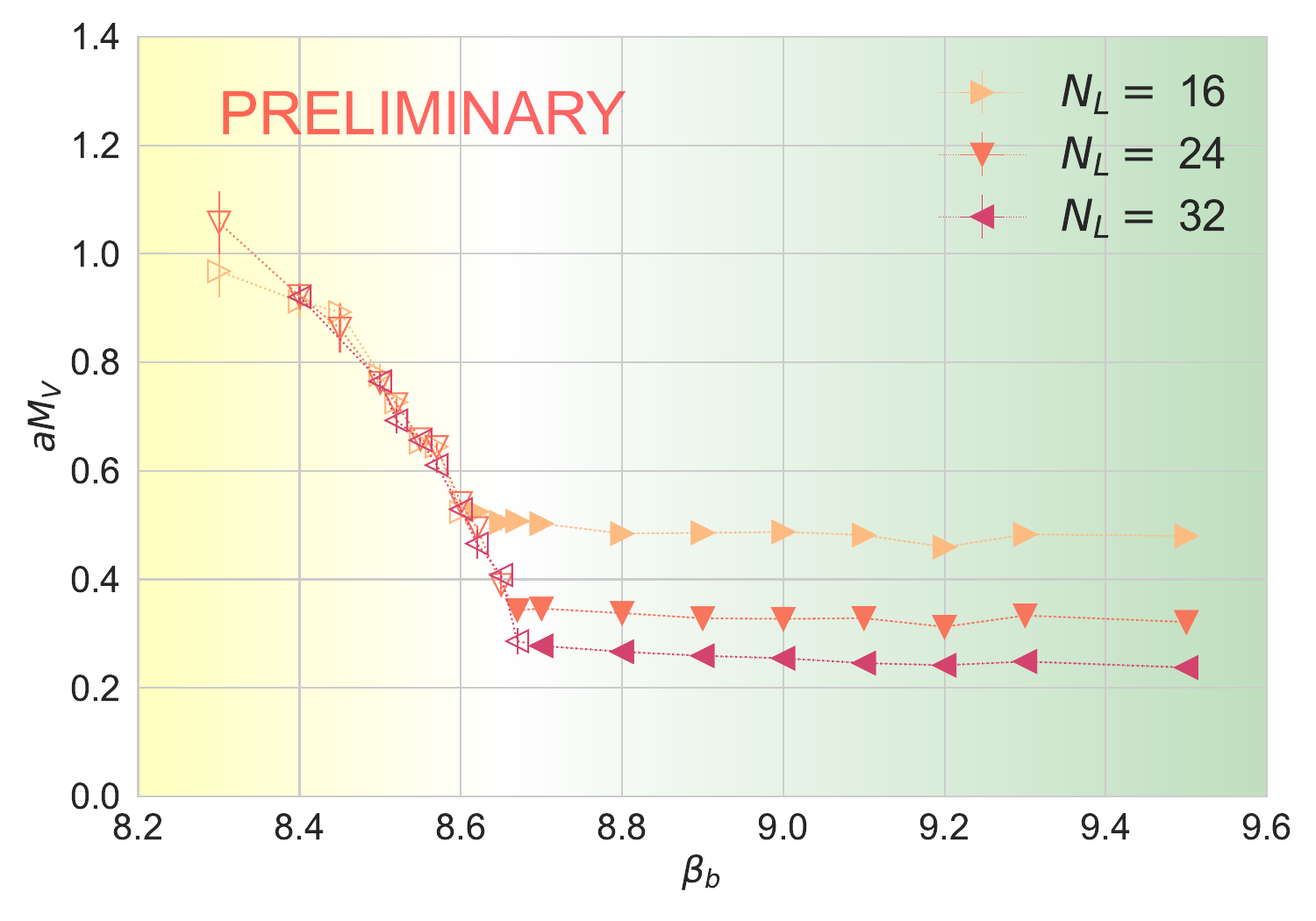}
  \caption{Mass of the pseudocalar (left) and vector meson (right) in lattice units vs.~the bare gauge coupling $\beta_b$. In the yellow-shaded strong coupling regime the volume dependence is weak and both states are massive, gapped. In the green-shaded weak-coupling region both states show strong volume dependence that is consistent with conformal hyperscaling as illustrated in Fig. \ref{Fig.LM-vs-beta}.  }
  \label{Fig.aM-vs-beta}
\end{figure}

We can confirm that the volume dependence at weak coupling is consistent with conformal hyperscaling, by showing in Fig.~\ref{Fig.LM-vs-beta} the same data as in Fig.~\ref{Fig.aM-vs-beta} multiplied by the spatial extent $N_L=L/a$ of the lattice. Now  at weak coupling both $L\,M_{PS}$ and $L\,M_V$ exhibit only minimal volume dependence, a sign of conformal hyperscaling.  In contrast, after the transition around $\beta_b\sim 8.65$ data corresponding to different volumes spread out.

Based on Figs.~\ref{Fig.aM-vs-beta} and ~\ref{Fig.LM-vs-beta} we conclude that the SU(3) 8-flavor system has two phases. The weak coupling phase appears conformal on our volumes.  This phase is chirally symmetric and shows conformal hyperscaling up to the phase transition. The strong coupling phase is also chirally symmetric, but its spectrum is gapped. Even the lightest connected meson state, $M_{PS}$, is massive, and mostly volume independent. This phase exhibits the properties of symmetric mass generation.

\begin{figure}[t]
  \includegraphics[width=0.49\textwidth]{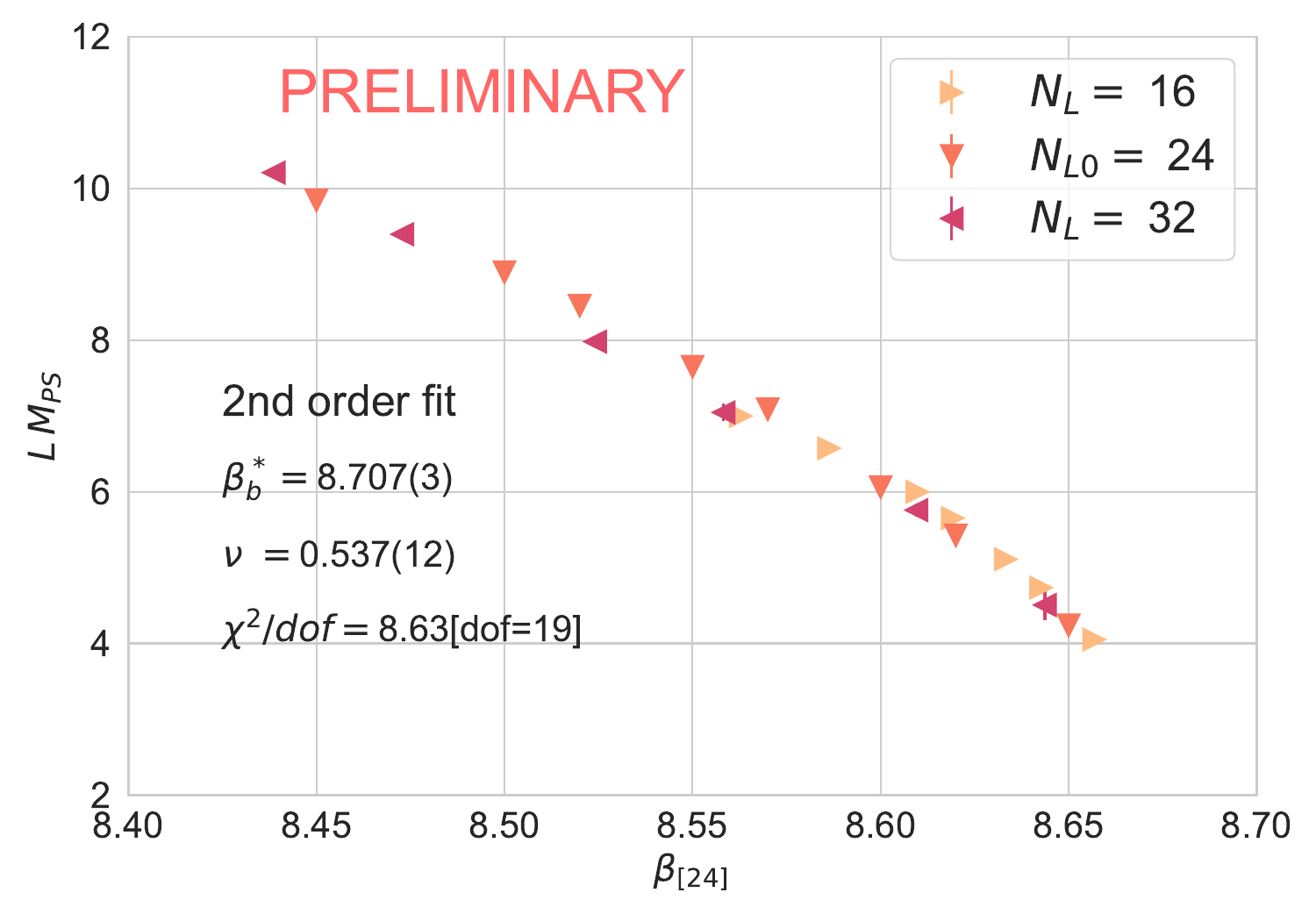}\hfill
  \includegraphics[width=0.49\textwidth]{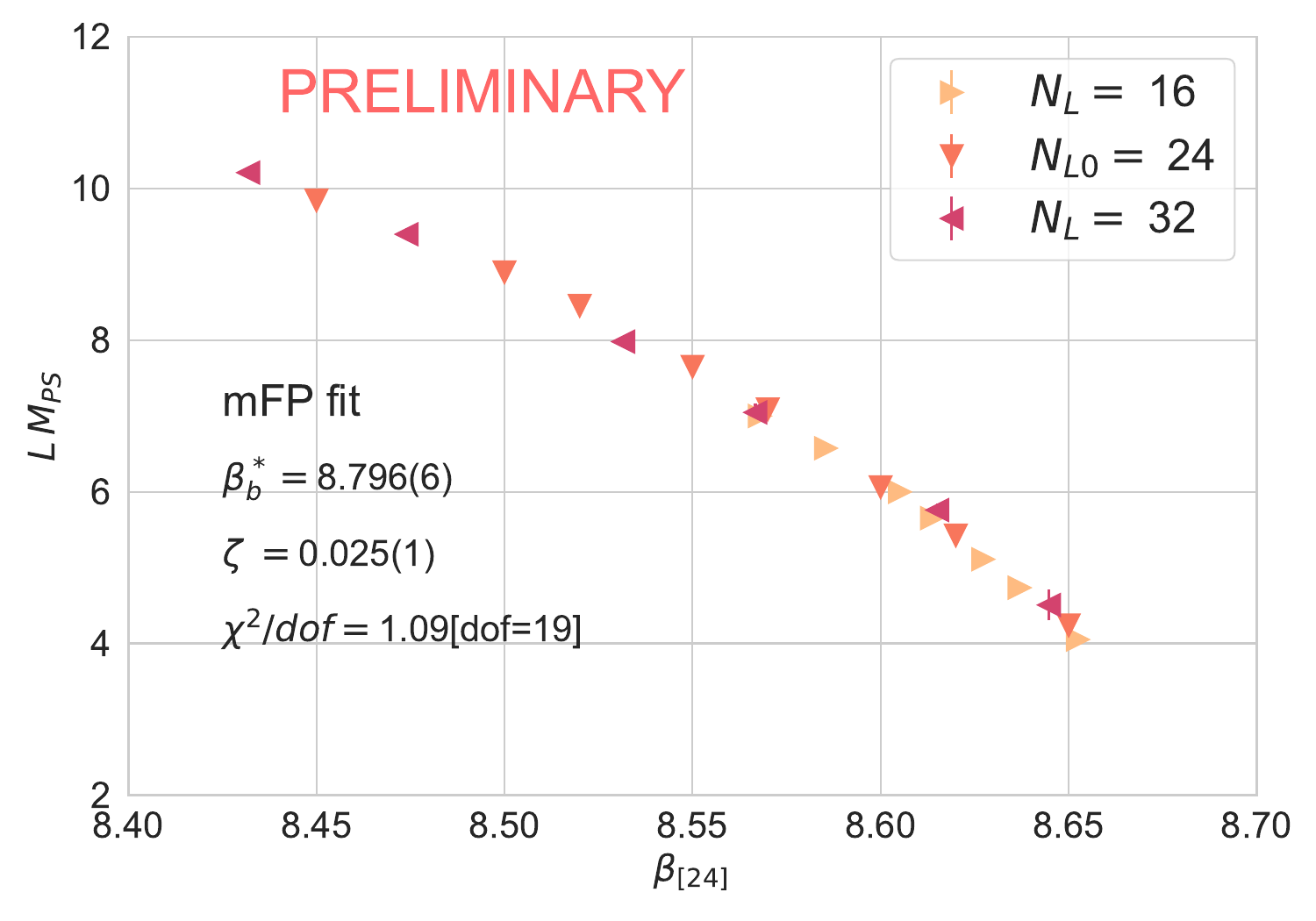}
  \caption{The result of the FSS curve collapse analysis assuming second order scaling (left panel) and merged fixed point (mFP) scaling (right panel). The results are presented in terms of the bare coupling of the reference volume $N_0=32$ as described in the text.}
  \label{Fig.FSS}
\end{figure}

Next we turn our attention to the phase transition separating these phases. We use the renormalization group invariant quantity $M_\text{PS}\,L$ to perform a finite size scaling (FSS) curve collapse analysis to determine the infinite volume critical coupling $\beta_b^\star$ and corresponding critical exponents. The explicit scaling form depends on the RG $\beta$ function. Specifically we consider three scenarios:
\begin{itemize}
\item {\it Second order phase transition}:  The RG $\beta$ function is linear near the critical point 
$\beta(g^2) =  - 1/\nu \,(g^2-g^2_\star)$, where $\nu$ is the critical exponent of the correlation length, $g^2=6/\beta_b$, and $g^2_\star=6/\beta_b^\star$.  The infinite volume lattice correlation length scales as $\xi \propto |\beta_b/\beta_* -1|^{1/\nu}$ and  RG invariant quantities in finite volume depend only on the scaling variable $\mathcal{X}= (\beta_b/\beta_* -1)N_L^{1/\nu}$. In particular, finite volume hadron masses scale as
\begin{equation}
M_\text{PS} \, L = f_\text{2nd}( \mathcal{X})     
\end{equation}
where $f_\text{2nd}$ is a unique function of the scaling variable $\mathcal{X}$. 
\item {\it Merged fixed point (mFP) transition}: The RG $\beta$ function has a quadratic zero near the critical point, $\beta(g^2) \propto (g^2-g^2_\star)^2$ . The infinite volume lattice correlation length is $\xi \propto e^{\left(\zeta /|\beta_b/\beta_* -1|\right)}$ and the finite volume scaling variable  is $\mathcal{X} =N_L e^{\left(-\zeta /|\beta_b/\beta_* -1|\right) }$. Finite volume hadron masses scale as
\begin{equation}
M_\text{PS} \, N_L = f_\text{mFP}( \mathcal{X}).     
\end{equation}
\item {\it First order phase transition}: While Wilsonian RG considerations do not apply directly to first order phase transitions, it is expected that first order scaling follows the second order scaling form with exponent $\nu=1/d$, with $d(=4)$ the dimension, in the regime where the lattice size is smaller/comparable to the correlation length.
\end{itemize}
We restrict the FSS curve collapse analysis to the strong coupling SMG phase, and follow the steps outlined in Ref.~\cite{Butt:2024kxi}. In Fig.~\ref{Fig.FSS} we present the results as the function of the bare coupling mapped to a reference volume $N_{L0}(=24)$. For second order scaling on volumes $N_L \ne N_{L0}$ we solve the equation $(\beta_b/\beta_* -1)N_L^{1/\nu} = (\beta_{[N_{L0}]}/\beta_* -1)N_{L0}^{1/\nu}$  to find $\beta_{[N_{L0}]}$ for given $(\beta_b,N_L)$, and similar for mFP scaling. This way the curve collapse analysis pulls all volumes to match the reference volume $N_{L0}$. 

The left panel of Fig. \ref{Fig.FSS} shows the result of the analysis using the second order, the right panel using the mFP scaling forms. While both fits show acceptable curve collapse, the mFP scaling fit has much smaller $\chi^2$ value. The fit quality becomes even worse for second order scaling when we systematically remove data points farther from the critical point. The quality and predicted parameters for the mFP fit remains largely unchanged under the same procedure. We conclude that our numerical data strongly prefer the merged fixed point scenario. However, even if the scaling followed the second order form, the predicted exponent $\nu \approx 0.537(12)$ is not consistent with a first order phase transition. The SMG phase thus exists in the continuum limit.

\begin{figure}[t]
  \includegraphics[width=0.47\textwidth]{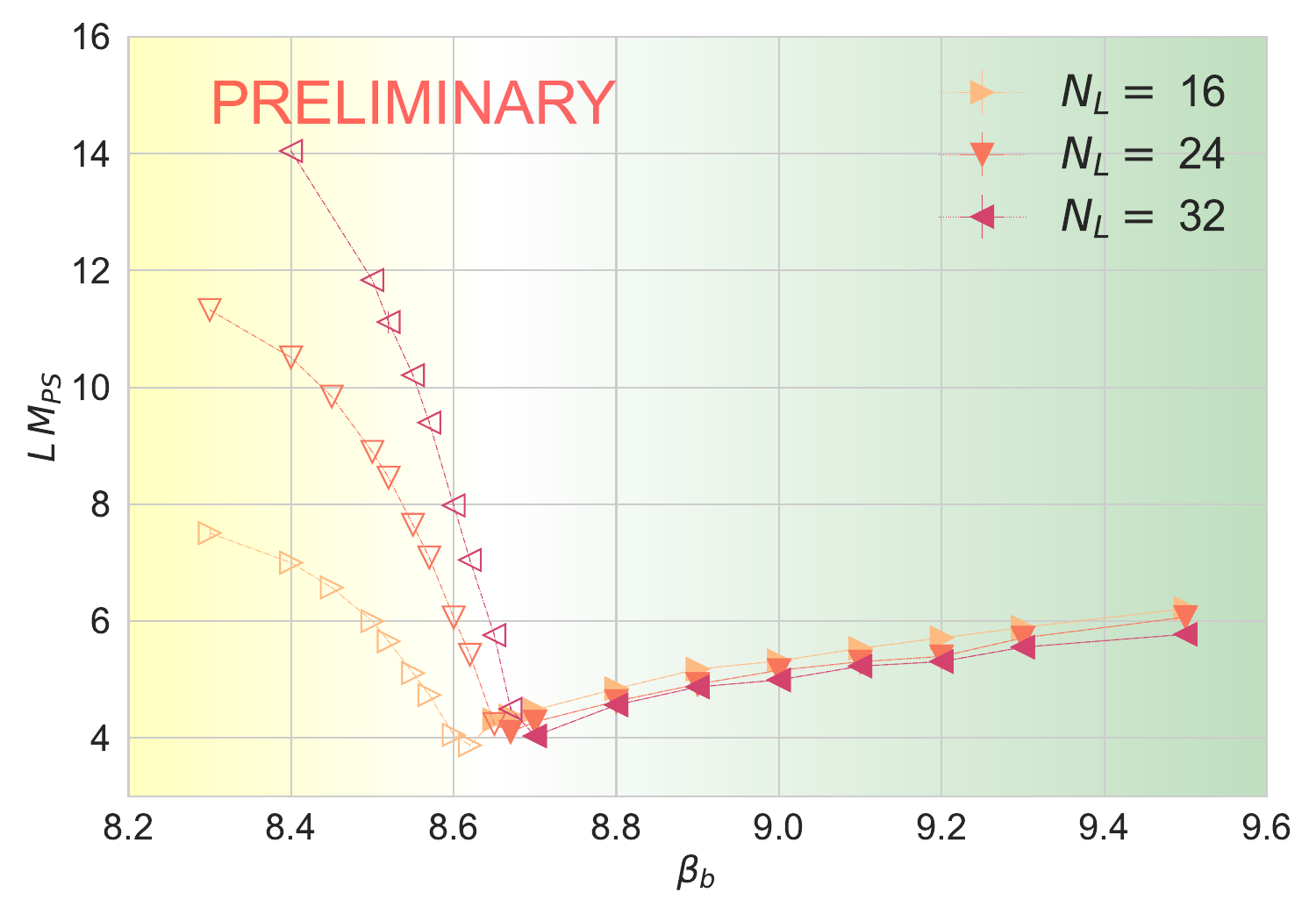}\hfill
  \includegraphics[width=0.47\textwidth]{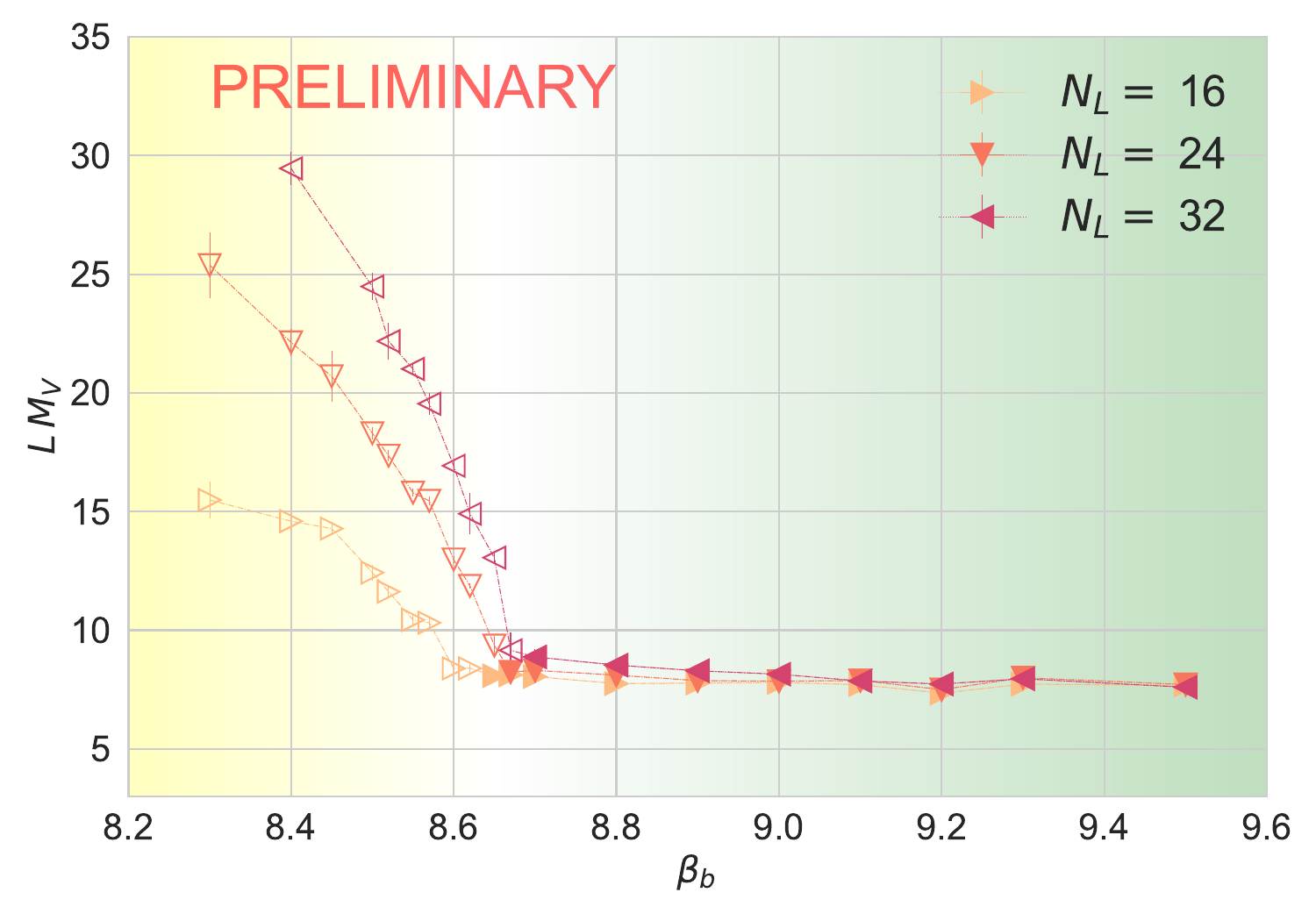}
  \caption{Same as Fig. \ref{Fig.aM-vs-beta} but for the quantity $L\,M_{PS}$ (left) and $L\,M_V$ (right). In the weak coupling, green-shaded regime the volume dependence largely disappears, consistent with conformal hyperscaling.
 }
  \label{Fig.LM-vs-beta}
\end{figure}

\section{Summary}
We report on our large scale investigations of the SU(3) gauge system with eight fundamental flavors. Using staggered fermions complemented with additional bosonic PV fields, we are able to push the simulations to strong renormalized coupling, where we observe that the system undergoes a phase transition. Studying the lowest lying meson masses we observe parity doubling over the entire simulation range. At weak coupling the different staggered tastes are almost degenerate but taste splitting grows as the coupling gets stronger. 

The phase transition occurs near $\beta_b=8.65$. Studying the volume dependence of the different hadronic states,  we find that in the weak coupling ($\beta_b\gtrsim 8.65$ data exhibit properties expected from conformal hyperscaling, whereas strong coupling data have properties matching an SMG phase. The system is chirally symmetric but gapped i.e.~even in the chiral limit we observe pseudoscalars with finite mass. 

The topological susceptibility also indicates that the vacuum structure of the strong coupling phase is different from the conformal or QCD-like vacuum. While the fermion Dirac operator does not have a zero mode, gradient flow smoothing reveals many instantons in the strong coupling, even though the simulations are performed in the massless chiral limit. The topological susceptibility is independent of the volume, implying that the instantons are part of the vacuum in the continuum limit.

To learn more about the nature of the phases and the phase transition itself, we perform finite size scaling analyses testing the hypotheses of a second order phase transition, a merged fixed point transition, and a first order phase transition. Our preliminary results from curve-collapse fits clearly prefer the merged fixed point scenario. This would imply that SU(3) with $N_f=8$ fundamental flavors is the onset of the conformal window.

In the future we plan to complement our simulations by using larger $48^3$ and, if viable,  $64^3$ volumes. This will help to boost confidence in our findings and in particular strengthen the FSS analysis as well as the hypothesis of conformal hyperscaling in the weak coupling regime. Moreover, we intend to perform simulations at finite mass, to allow for extrapolations to the zero mass limit and that way add information on the SMG phase. 

\acknowledgments
AH.~acknowledges support from DOE grant DE-SC0010005.

The numerical simulations were performed using the Quantum EXpressions (QEX) code  \cite{Osborn:2017aci,Jin:2016ioq} for gauge field generation, Qlua \cite{Pochinsky:2008zz,qlua} for gradient flow measurements and hadronic spectra are calculated using MILC \cite{MILC}. We thank in particular James Osborn, Xiaoyong Jin, and Curtis T. Peterson for their assistance in starting this project.

Computational resources provided by the USQCD Collaboration, funded by the Office of Science of the U.S.~Department of Energy, Lawrence Livermore National Laboratory (LLNL), Boston University (BU) computers at the MGHPCC, in part funded by the National Science Foundation (award No.~OCI-1229059), and the OMNI cluster of the University of Siegen were used. We thank the LLNL Multiprogrammatic and Institutional Computing program for Grand Challenge supercomputing allocations and thank Amitoj Singh and his team for early science time on the 24s cluster at Jefferson Lab.

This document was prepared using the resources of the USQCD Collaboration at
the Fermi National Accelerator Laboratory (Fermilab), a U.S.~Department of
Energy (DOE), Office of Science, HEP User Facility. Fermilab is managed by Fermi Research
Alliance, LLC (FRA), acting under Contract No.~DE-AC02-07CH11359.
\bibliography{./BSM}
\bibliographystyle{JHEP-arXiv}


\end{document}